\documentclass[preprint,12pt]{elsarticle}

\usepackage[T1]{fontenc}

\usepackage{amsmath,amsfonts,amssymb,amsthm}
\usepackage{latexsym}

\usepackage{graphicx}

\usepackage{booktabs}
\usepackage{array}
\usepackage{tabularx}

\DeclareMathOperator{\Ramp}{Ramp}
\DeclareMathOperator{\var}{var}
\DeclareMathOperator{\Fr}{Fr}

\newcolumntype{C}{>{\centering\arraybackslash$}X<{$}}

\allowdisplaybreaks[1]

\journal{Physica A}

\begin{document}

\begin{frontmatter}

\title{A Dynamical Model for Forecasting Operational Losses}

\author[uniba,infn_ba]{M. Bardoscia \corref{mail}}
\cortext[mail]{Corresponding author mail: \texttt{marco.bardoscia@ba.infn.it}}
\author[uniba,infn_ba]{R. Bellotti}
\address[uniba]{Dipartimento Interateneo di Fisica ``\emph{M.Merlin}'', Universit\`a degli Studi di Bari e Politecnico di Bari, via Amendola 173, I-70126 Bari, Italy}
\address[infn_ba]{Istituto Nazionale di Fisica Nucleare, Sezione di Bari, \\ via Amendola 173, I-70126 Bari, Italy}

\begin{abstract}
A novel dynamical model for the study of operational risk in banks and suitable for the calculation of the Value at Risk (VaR) is proposed. The equation of motion takes into account the interactions among different bank's processes, the spontaneous generation of losses via a noise term and the efforts made by the bank to avoid their occurrence. Since the model is very general, it can be tailored on the internal organizational structure of a specific bank by estimating some of its parameters from historical operational losses. The model is exactly solved in the case in which there are no causal loops in the matrix of couplings and it is shown how the solution can be exploited to estimate also the parameters of the noise. The forecasting power of the model is investigated by using a fraction $f$ of simulated data to estimate the parameters, showing that for $f = 0.75$ the VaR can be forecast with an error $\simeq 10^{-3}$.
\end{abstract}


\end{frontmatter}

\section{Introduction} \label{sec:intro}
The methods developed in the context of statistical mechanics and, more in general, in the study of complex systems have found in the last years broad application in many different scientific fields. Economic sciences particularly benefited from interdisciplinary approaches and borrowed some crucial ideas, powerful tools and techniques \cite{mantegna-stanley} from those fields. However, these efforts have been devoted almost exclusively to the study of the financial risk \cite{bouchaud-potters}, and only more recently also other typologies of risk \cite{voit} as the operational risk \cite{mcneil-frey-embrechts, cruz} are gaining more and more attention.

Operational risk is ``the risk of [money] loss [in banks] resulting from inadequate or failed internal processes, people and systems or from external events'' \cite{basel}, including legal risk, but excluding strategic and reputation linked risks. Let us make an example to clarify the dynamics underlying the generation of operational losses; suppose that a material damage in the system that controls and authorizes the transactions occurs and is discovered at the time $t_1$, but repaired only later at the time $t_2$; a loss equal to the amount of money needed to repair the damage is generated at the time $t_1$ in the process of machinery servicing, but the failure has likely generated losses delayed up to the time $t_2$, because some transactions have failed or have been wrongly authorized. This example shows that the different processes may be strongly correlated, and that their typical correlations extend over time.

The primary goal of the management of operational risk is to determine the capital charge that the bank has to put aside (e.g.\ every year) to cover the operational losses. The New Basel Capital Accord \cite{basel} roughly proposes to set this capital to the $15\%$ of the bank's gross income, or to consider the gross income per business line and weight each one with a coefficient ranging from $12\%$ to $18\%$; we will call these approaches ``macroscopic'', since they assume that the capital requirement does not depend on the details of the internal structure of the bank, but only on its size. However, the basic assumption of these approaches seems not to be realistic and they do not provide any insight on the mechanisms underlying the generation of losses, not allowing any practice aimed to foresee or reduce the future losses.

The New Basel Capital Accord also envisages that each bank is free to develop its own approach to the evaluation of the capital requirement as long as it satisfies some general requirements. One possible approach is the ``microscopic'' one in which one tries to take into account all the fundamental mechanisms involved in the generation of operational losses; it is clear that in this framework one should deal not only with variables representing operational losses but with heterogeneous variables which strongly depend on the particular mechanism under examination. Although these approaches are more solidly founded than the macroscopic ones, they suffer from practical drawbacks; firstly, it is extremely difficult to introduce a dynamical model that couples all the microscopic variables in a realistic way: for this reason some attempts in this sense \cite{cowell-verral-yoon} have been done using bayesian networks, which allow to model statistical correlations among heterogeneous variables, but do not allow to follow the evolution of losses in time. Secondly, the implementation of such an approach inside a bank requires that all the relevant variables should be monitored reliably with a certain frequency, which can be an extremely resource demanding task, especially for a small or medium sized bank.

An alternative approach may be to provide a description of loss events based on an effective model consisting of much fewer degrees of freedom than a microscopic one, but still able to distinguish the internal structure of a bank; it is natural to call such an approach ``mesoscopic'', since one can think that the effective variables are obtained by integrating out the details contained in the microscopic ones. In the non-macroscopic approaches the capital requirement is usually identified with the Value-at-Risk (VaR) over one year and with $99.9\%$ level of confidence, i.e.\ the $99.9$ percentile of the yearly loss distribution; this implies that the probability of registering a loss being greater than the value of the VaR in one year is equal to $0.001$ or, equivalently, that such a loss may occur on average every $1000$ years. The most widely used non-macroscopic approach is mesoscopic: the Loss Distribution Approach (LDA) \cite{frachot-moudoulaud-roncalli, aue-kalkbrener, shevchenko} classifies the loss events in $8$ business lines (sectors of activity of the bank) and $7$ event types (causes of the loss) and identifies the relevant variables with the number of losses occurred during a certain time horizon (frequency) and with the amount of a single loss (severity) for each of the $56$ couples (business line, event type); the LDA makes use of the distributions of frequency and severity to derive the loss distribution over the given time horizon, usually assuming that frequency and severity for each process are independent random variables, and thus failing in capturing the correlations among different couples. There are some proposals of how to take into account the correlations in the context of LDA, e.g.\ among frequency of different couples \cite{aue-kalkbrener, powojowski-reynolds-tuenter, bee}, frequency and severity of the same couple \cite{neil-fenton-tailor}, frequency and severity of different couples \cite{frachot-roncalli-salomon, boecker-klueppelberg} or aggregate losses of different couples \cite{frachot-roncalli-salomon, gourier-farkas-abbate}, or in different frameworks \cite{bonafede-giudici, bardoscia-et-al}, but no one has gained a general consensus. It is worth pointing out that it is very unlikely that mechanism of loss production involves frequency and severity, which should be simply considered statistical tools to model the loss distribution over the given time horizon: for this reason it is not obvious how to incorporate the time dependence into the LDA framework.

A different possible mesoscopic approach consists in assuming that the effective variables are the degrees of freedom of a dynamical system and postulating an effective equation of motion \cite{leippold-vanini}. The model should be sufficiently general to explain the dynamics of loss production in all the banks, but flexible enough to adapt to the particular internal structure of a specific bank, for example by properly tuning the parameters appearing in the equation of motion. Once the parameters have been estimated, the advantage of a dynamical approach is immediately evident: one may follow the production of the losses during time and thus may be able to make predictions about the evolution of losses. In the approach presented in this paper the effective variables are the amount of losses registered at a certain time in a certain process; processes are categories in which losses are classified and depend on the specific structure of a bank; examples are material damage, failed transaction or fraud. The equation of motion includes two general mechanisms for the generation of losses in a process: the interactions with other processes and the spontaneous generation due to a random noise; the possibility that the bank invests money to avoid the occurrence of losses is also taken into account. As the equation of motion contains a noise term, the loss distribution will naturally arise considering several realization of the noise: therefore the VaR can be still taken as a measure of the capital requirement. Since the different-time correlations play a crucial role, the interaction term is non-local in time. Let us point out that, even in the case in which the microscopic dynamics of the system was local in time, it is perfectly reasonable to assume an effective dynamics that is non-local in time: it is a well known result \cite{zwanzig} that the reduced dynamical system obtained integrating out some degrees of freedom from a dynamical system with equation of motion local in time is in general characterized by an equation of motion which is non-local in time. From this point of view, the non-locality in the equation of motion is justified a priori, basing on very general considerations, rather than a posteriori, basing on some peculiar features of the loss distributions. 

A dynamical model for operational risk has been already proposed in \cite{kuhn-neu-2003, anand-kuhn}, and also applied to the study of credit risk \cite{kuhn-neu-2004}. There are some important differences between the approach in \cite{kuhn-neu-2003, anand-kuhn} and the one proposed in this paper. Firstly, while the dynamics in \cite{kuhn-neu-2003, anand-kuhn} is local in time, the one proposed here is not, for the reasons that we have just discussed. Secondly, as explained in Section\ \ref{sec:alt_model}, the dynamics proposed here allows the estimation of the parameters of the noise. In addition, even if it is possible to show that the proposed dynamics is equivalent to a dynamical generalization of the LDA \cite{kuhn-neu-2003} (see Section\ \ref{sec:alt_model}), it is possible to interpret all the terms in the equation of motion as general mechanisms which are responsible for the generation of operational losses. From this point of view, it is natural to build a dynamics which directly involves the amount of losses registered in each process. Such an approach has been introduced in \cite{bardoscia-bellotti}, where also a detailed comparison with the framework of frequency and severity is made. The methodological advantage is that one has not to make direct assumptions on the shape of the loss distribution, but only on the basic mechanisms that generate the losses: in a sense, the features of the loss distribution \emph{emerge} from those mechanisms. This is a fundamental difference with respect to \cite{kuhn-neu-2003, anand-kuhn}, where the form of both frequency and severity distributions are imposed a priori.

The paper is organized as follows: in Section\ \ref{sec:model} the model is introduced and in Section\ \ref{sec:solutions} it is shown that under some hypothesis it can be exactly solved; in Section\ \ref{sec:learning} it is illustrated how some parameters of the model can be estimated from real data; in Section\ \ref{sec:tests} the proposed procedure to estimate the parameters is validated by means of simulated data; moreover, the capability of model to forecast future operational losses is tested by estimating the parameters only from a fraction of simulated data and comparing the forecasts made by the model with the remaining part; in Section\ \ref{sec:alt_model} the model is compared with an alternate one, while in Section\ \ref{sec:outro} some conclusions are drawn.

\section{The Model} \label{sec:model}
The model consists of $N$ positive real variables $l_i(t)$ that represent the amount of loss (in some currency) registered in the process $i$ at the time $t$ and that evolve by means of a discrete time equation of motion. The variables are coupled through the matrix $J$ which in general is not symmetric: $J_{ij} \neq 0$ means that $l_i$ is influenced by $l_j$ and not vice versa; the equation of motion is non-local in time in the sense that, if $J_{ij} \neq 0$, $l_i(t)$ depends on $l_j(t-1), \ldots, l_j(t-t_{ij}^*)$ which are the values that $l_j$ takes in the past $t_{ij}^*$ time steps; $t_{ij}^*$ can thus be thought as an asymmetric time of correlation between the variables $l_j$ and $l_i$. The equation of motion is:
\begin{equation} 	\label{eq:motion}
	l_i(t) = \Ramp \left( \sum_{j=1}^N J_{ij} C_{ij}(t) + \theta_i + \xi_i(t) \right) \, ,
\end{equation}
where the ramp function:
\begin{equation*}
	\Ramp(x) = 
	\begin{cases}
		x & \text{for} \;\; x > 0 \\
		0 & \text{for} \;\; x \leq 0
	\end{cases}
\end{equation*}
ensures that $l_i(t) \in \mathbb{R^+}$, $\forall \, t$. The positive terms in the argument of the ramp function in \eqref{eq:motion} tend to generate a loss, while the negative terms tend to avoid the occurrence of a loss. The presence of the ramp function in \eqref{eq:motion} excludes the possibility of negative losses which could be interpreted as reserves of money put aside to automatically lower future losses.

$C_{ij}(t)$ simply counts the number of $l_j(t)$ greater than zero in the time interval $[t-t_{ij}^*, \, t-1]$:
\begin{equation} \label{eq:trigger}
	C_{ij}(t) = \sum_{1 \leq s \leq t_{ij}^*} \Theta \left[ l_j(t-s) \right] \, ,
\end{equation}
where $\Theta$ is the Heaviside function. Eq.\ \eqref{eq:trigger} implies that $C_{ij}(t) \in \{0, 1, \ldots, \, t_{ij}^*\}$ and the coupling term in \eqref{eq:motion} can assume only the values $0, J_{ij}, 2 J_{ij}\ldots, t_{ij}^* J_{ij}$, so that, if $J_{ij} \neq 0$, $l_i(t)$ does not depend on the values of $l_j(t-s)$, but only on the number of times in which $l_j(t-s)$ is greater than zero for $s \in [t-t_{ij}^*, \, t-1]$. This means that, if $J_{ij} > 0$, each loss occurred in the process $j$ between the time steps $t - t_{ij}^*$ and $t - 1$ generates a \emph{potential} loss of amount $J_{ij}$ in the process $i$ at time $t$; on the other hand $J_{ij} < 0$ means that a loss in the process $j$ may help the process $i$ to function properly. Such an interaction term implies the following approximation: a potential loss generated by other losses does not depend on their amount, but only on their number within a certain maximum correlation time. The non-locality in time of \eqref{eq:trigger} is crucial to take into account the different-time correlations, as pointed out in Section\ \ref{sec:intro}. Let us incidentally notice that \eqref{eq:motion} requires an initial condition consisting of a number of time steps equal to the maximum of $t_{ij}^*$.

The inhomogeneous external field $\theta_i$ has two very different interpretations depending on its sign; a field term $\theta_i < 0$ can be interpreted as the effort (investment) made by the bank to avoid the occurrence of losses in the process $i$: in fact the sum of the interaction term and $\xi_i(t)$ has to be greater than $|\theta_i|$ to effectively produce a loss. In this scenario the fact that $\theta_i$ does not depend on time implies that the amount of money (per unit of time) to invest on each process is chosen a priori and kept fixed for a long period of time, rather than dynamically adjusted ``on the fly''. A field term $\theta_i > 0$ could be interpreted as a pathological tendency of the process $i$ to produce losses at every time step and thus is undesirable in this context.

$\xi_i(t)$ is a random noise  $\delta$-correlated in time that accounts for spontaneously generated losses, i.e.\ losses that are not caused by the occurrence of other losses; this interpretation implies that it must have a positive support. As discussed in detail in Section\ \ref{sec:solutions}, the analytical results that will be obtained are very general, in the sense that they can be easily extended to different distributions of the noise, provided it satisfies some very general hypotheses. To fix the ideas we choose $\xi_i(t)$ to have an exponential distribution:
\begin{subequations} \label{eq:noise}
	\begin{equation}
		\rho( \xi_i ) = \lambda_i e^{-\lambda_i \xi_i}
	\end{equation}
	\begin{equation}
		\langle \xi_i(t) \rangle = \frac{1}{\lambda_i}
	\end{equation}
	\begin{equation}
		\langle \xi_i(t) \xi_j(s) \rangle = \frac{1}{\lambda_i}  \delta_{i,j} \delta_{t,s} \, ;
	\end{equation}
\end{subequations}
the rationale behind such a choice is the following: as it can be intuitively argued, spontaneous losses (like those caused by human errors, or machine failures) are relatively rare events: such a behavior can be obtained by setting $\theta_i < 0$ and $| \theta_i | > 1 / \lambda_i$ since the chosen distribution is exponential and the majority ($\simeq 63\%$) of the \emph{potential} losses generated by the noise are smaller than its mean value $1 / \lambda_i$. Because of the presence of noise in \eqref{eq:motion} $l_i(t)$ is a random variable; from this point of view we say that the model can be exactly solved if all the moments of the probability distribution of $l_i(t)$ can be calculated.

The crucial quantity for the study of operational risk is the cumulative loss up to the time $t$:
\begin{equation} \label{eq:zeta}
	z_i(t) = \sum_{s \leq t} l_i(s) \, ,
\end{equation}
which can be taken as an approximated indicator of the capital that should be put aside to face operational risk over a time horizon $t$.

\section{Model Solutions} \label{sec:solutions}
In this section it will be shown that, if the structure of the coupling matrix $J$ satisfies some peculiar hypotheses, the model can be exactly solved in the sense specified in Section\ \ref{sec:model} and the asymptotic behaviour of $z_i(t)$ can be determined.

We give some preliminary definitions: a process $i$ is said to be \emph{influenced} by a process $j$ if $J_{ij} \neq 0$; a process $i$ is said to be \emph{free} if it is not influenced by any process (including itself), i.e.\ $J_{ij} = 0$, $\forall \, j$. These definitions are coherent with the mechanism of the interaction term in \eqref{eq:motion}: in fact a loss occurred in the process $j$ may cause a loss in the process $i$ only if $J_{ij} \neq 0$. The hypothesis on the structure of $J$ can be stated in the following way: let us associate to each process a node in a graph and, if the process $i$ is influenced by the process $j$, let us draw a directed edge from the node $j$ to the node $i$; the graph obtained considering only the nodes influencing directly or indirectly the node $i$, together with the node $i$ itself will be called the subgraph associated to the process $i$; if the subgraph associated to the process $i$ is a directed acyclic graph, i.e.\ if the edges in the graph do not form any closed loop (see \cite{thulasiraman-swamy} for basic definitions about graphs), all the moments of the distribution of $\langle l_i(t) \rangle$ can be exactly calculated. In such a case we say that the subgraph associated to the process $i$ \emph{has no causal loops}; the meaning of this definition can be understood considering a graph with a loop like $i \rightarrow j \rightarrow i$: in such a case the losses occurred in the process $i$ may cause other losses in the process $j$, which in turn may cause other losses in the process $i$, resulting in a causal loop. If the whole graph associated with the coupling matrix $J$ is a directed acyclic graph we say that the matrix $J$ has no causal loops: in such a case the subgraphs associated to all the processes have no causal loops, and therefore the model can be exactly solved. We remark that the absence of causal loops is a commonly accepted hypothesis, e.g.\ in the context of other tools which are widely used to take into account the correlations among different process, like bayesian networks \cite{cowell-verral-yoon, neil-fenton-tailor, bonafede-giudici, bardoscia-et-al}.

Only the cases relative to the two simplest subgraphs will be treated here, deferring a more general discussion to the Appendix. Let us start with a free process $i$, i.e.\ the subgraph associated to the process $i$ is just a node with no incident edges. In this case the random variable $l_i(t)$ is independent from $l_j(t')$, $\forall \, j, \, t'$ and the $n$-th moment of its distribution is simply the average of $l_i^n(t)$ over the noise, i.e.\ the average over the random variable $\xi_i(t)$ (we will use $\tilde{d \xi_i}(t)$ as a shorthand for $\rho(\xi_i) \, d \xi_i(t)$):
\begin{equation}
	\begin{split}
		\langle l_i^n(t) \rangle & 
		= \int_{0}^{\infty} l_i^n(t) \, \tilde{d \xi_i}(t) \\
		& = \int_{0}^{\infty} \Ramp \left[ \theta_i + \xi_i(t) \right]^n \, \tilde{d \xi_i}(t) \, ;
	\end{split}
\end{equation}	
defining
\begin{equation} \label{eq:free_mom_def}
	m_i^{(n) F}(x) \equiv \int_{0}^{\infty} \Ramp \left[ x + \xi_i(t) \right]^n \, \tilde{d \xi_i}(t)
\end{equation}
we have:
\begin{equation} \label{eq:free_1}
	\langle l_i(t) \rangle = m_i^F(\theta_i) = 
		\begin{cases}
			\frac{e^{\lambda_i \theta_i}}{\lambda_i} & \text{if} \;\; \theta_i < 0 \\
				\theta_i + \frac{1}{\lambda_i} & \text{if} \;\; \theta_i \geq 0
		\end{cases} \, ,
\end{equation}
where $m_i^F$ has been used (and will be used in the following) as a shorthand for $m_i^{(1) F}$. The mean of $l_i^2(t)$ can be analogously calculated:
\begin{equation} \label{eq:free_1_bis}
	\langle l_i^2(t) \rangle = m_i^{(2) F}(\theta_i) = 
		\begin{cases}
			\frac{2 \, e^{\lambda_i \theta_i}}{\lambda_i^2} & \text{if} \;\; \theta_i < 0 \\
			\theta_i^2 + \frac{2 \, \theta_i}{\lambda_i} + \frac{2}{\lambda_i^2} & \text{if} \;\; \theta_i \geq 0
		\end{cases} \, ,
\end{equation}
so that the variance of $l_i(t)$ is:
\begin{equation} \label{eq:free_2}
	\begin{split}
		\var l_i(t) & = \langle l_i^2(t) \rangle - \langle l_i(t) \rangle^2 \\
		& =	\begin{cases}
				\frac{e^{\lambda_i \theta_i}}{\lambda_i^2} (2 - e^{\lambda_i \theta_i}) & \text{if} \;\; \theta_i < 0 \\
				\frac{1}{\lambda_i^2} & \text{if} \;\; \theta_i \geq 0
			\end{cases} \, .
	\end{split}
\end{equation}
As expected for a free process, $\langle l_i(t) \rangle$ and $\var l_i(t)$ do not depend on time, and all the moments of the probability distribution of $l_i(t)$ do not as well; $z_i(t)$ is thus the sum of $t$ independent and identically distributed (i.i.d.)\ variables with finite variance and, by means of the central limit theorem, for sufficiently large $t$ it has a Gaussian distribution with mean and variance:
\begin{subequations} \label{eq:zeta_analytical}
	\begin{equation}
		\langle z_i(t) \rangle = t \, \langle l_i(t) \rangle
	\end{equation}
	\begin{equation}
		\var z_i(t) = t \, \var l_i(t) \, .
	\end{equation}
\end{subequations}

The next step is to repeat the calculation of \eqref{eq:free_1} and \eqref{eq:free_2} for the process $i$ in the case in which it is influenced only by a single process $j$ and the process $j$ is a free process ($i \leftarrow j$). Since in this case $l_i(t)$ depends through $C_{ij}(t)$ only on $l_j(t-1), \ldots, l_j(t-t_{ij}^*)$, the average over the noise equals to the average over the random variables $\xi_i(t), \, \xi_j(t-1), \ldots, \xi_j(t-t_{ij}^*)$:
\begin{equation} \label{eq:one_leaf_1}
	\langle l_i(t) \rangle = \int_0^{\infty} \Ramp \left[ J_{ij} C_{ij}(t) + \theta_i + \xi_i(t) \right]
	\prod_{1 \leq s \leq t_{ij}^*} \tilde{d \xi_j}(t-s) \; \tilde{d \xi_i}(t) \, ;
\end{equation}
let us observe that the domain of integration of the variables $\xi_j(t-1), \ldots, \xi_j(t-t_{ij}^*)$ can be divided in subsets obtained by fixing the value of $C_{ij}(t)$; since the events $C_{ij}(t)=0, \ldots, C_{ij}(t)=t_{ij}^*$ are mutually exclusive and cover the entire domain of integration:
\begin{equation} \label{eq:one_leaf_2}
	\prod_{1 \leq s \leq t_{ij}^*} \int_0^{\infty} \tilde{d \xi_j}(t-s) = \sum_{c=0}^{t_{ij}^*} \int_{C_{ij}(t) = c} \prod_{1 \leq s \leq t_{ij}^*} \tilde{d \xi_j}(t-s) \, .
\end{equation}
Each term in the summation on the right hand side of \eqref{eq:one_leaf_2} is simply the probability that $C_{ij}(t)=c$, i.e.\ the probability that $c$ elements in the set $\{ l_j(t-1), \ldots, l_j(t-t_{ij}^*) \}$ are greater than zero and $t_{ij}^* - c$ elements are less than or equal to zero; since the process $j$ is free, the probability that $l_j(t) > 0$ is easily calculated:
\begin{equation}
	\begin{split}
 		\Pr \left[ l_i(t) > 0 \right] & = 
		\int_{0}^{\infty} \Theta \left[ l_i(t) \right] \, \tilde{d \xi_i}(t) \\
		& = \int_{0}^{\infty} \Theta \left[ \theta_i + \xi_i(t) \right] \, \tilde{d \xi_i}(t) \\
	\end{split}
\end{equation}
defining
\begin{equation} \label{eq:free_p_def}
	p_i^F(x) \equiv \int_{0}^{\infty} \Theta \left[ x + \xi_i(t) \right] \, \tilde{d \xi_i}(t)
\end{equation}
we have:
\begin{equation} \label{eq:free_3}
	\Pr \left[ l_i(t) > 0 \right] = p_i^F(\theta_i) =
		\begin{cases}
			e^{\lambda_i \theta_i} & \text{if} \;\; \theta_i < 0 \\
			1 & \text{if} \;\; \theta_i \geq 0
		\end{cases}
		\, ,
\end{equation}
that yields:
\begin{equation} \label{eq:one_leaf_3}
	\int_{C_{ij}(t) = c} \prod_{1 \leq s \leq t_{ij}^*} \tilde{d \xi_j}(t-s)
	= \binom{t_{ij}^*}{c} \left[ p_j^F(\theta_j) \right]^c \left[ 1 - p_j^F(\theta_j) \right]^{t_{ij}^* - c} \, .
\end{equation}
Using \eqref{eq:one_leaf_3} and \eqref{eq:free_mom_def}, \eqref{eq:one_leaf_1} becomes:
\begin{equation} \label{eq:one_leaf_4}
	\begin{split}
		\langle l_i(t) \rangle & =	
		\sum_{c=0}^{t_{ij}^*} \int_{C_{ij}(t) = c} \prod_{1 \leq s \leq t_{ij}^*} \tilde{d \xi_j}(t-s) 
		\int_0^{\infty} \Ramp \left[ c J_{ij} + \theta_i + \xi_i(t) \right] \, \tilde{d \xi_i}(t) \\
		& = \sum_{c=0}^{t_{ij}^*}  \binom{t_{ij}^*}{c} \left[ p_j^F(\theta_j) \right]^c \left[ 1 - p_j^F(\theta_j) \right]^{t_{ij}^* - c} \, 
		m_i^F(c J_{ij} + \theta_i) \, .
	\end{split}
\end{equation}
The same line of reasoning leading from \eqref{eq:one_leaf_1} to \eqref{eq:one_leaf_4} can be followed to calculate the variance:
\begin{equation} \label{eq:one_leaf_5}
	\var l_i(t) = \left[ \sum_{c=0}^{t_{ij}^*}  \binom{t_{ij}^*}{c} \left[ p_j^F(\theta_j) \right]^c \left[ 1 - p_j^F(\theta_j) \right]^{t_{ij}^* - c} \, m_i^{(2)F}(c J_{ij} + \theta_i) \right] - \langle l_i(t) \rangle^2 \, 
\end{equation}
or any moment of the distribution of $l_i(t)$. Even in this case $z_i(t)$ is the sum i.i.d.\ variables with finite variance and thus \eqref{eq:zeta_analytical} is also valid; in the Appendix it is shown that \eqref{eq:zeta_analytical} still holds in the more general case in which the coupling matrix $J$ has no causal loops; as a consequence, the non-locality of the equation of motion alone is not sufficient to modify the shape of the cumulative loss distribution. Actually \eqref{eq:zeta_analytical} has a crucial importance: while, at least in principle, it is possible to think at an extension of the technique used here and in the Appendix to calculate the moments of $l_i(t)$ also in the case in which the matrix $J$ has causal loops, the random variables $l_i(t)$ for different values of $t$ would be neither independent nor identically distributed in that case, and \eqref{eq:zeta_analytical} would not hold anymore. Determining the moments of $z_i(t)$ by the explicit calculation of the moments of $l_i(s)$, $\forall \, s \leq t$ is also hopeless, since it would become exponentially complex in $t$, as shown in the Appendix. It is worth noting that \eqref{eq:one_leaf_4}, \eqref{eq:one_leaf_5} and their analogous in the Appendix reduce the calculation of the mean and the variance of $l_i(t)$ to the calculation of $m_i^F$ and $m_i^{(2)F}$ and $p_i^F$, which are the only quantities that depend on the particular distribution of the noise. This means that those expressions may be easily generalized to a different distribution of the noise simply by recalculating $m_i^F$, $m_i^{(2)F}$ and $p_i^F$ from \eqref{eq:free_mom_def} and \eqref{eq:free_p_def}, provided that the corresponding integrals converge; in particular, while $p_i^F$ is always finite, it can be easily shown that $m_i^F$ ($m_i^{(2)F}$) is finite if and only if the mean (second central moment) of the noise is finite. However, if $m_i^{(2)F}$ diverges, the central limit theorem does not apply and the distribution of $z_i(t)$ at large $t$ is not Gaussian.

\section{Parameters Estimation} \label{sec:learning}
In this section a scheme for estimating the parameters of the model from real data will be presented. In the more general case $\vec{\theta}$ and $J$ can be estimated, but the parameters $\vec{\lambda}$ of the noise must be known a priori. If the graph associated to the matrix $J$ is known and has no loops, i.e.\ if according to the definition given in Section\ \ref{sec:solutions} the matrix $J$ has no causal loops, the model can be integrated and the additional constraint imposed by the exact solution can be exploited to estimate also $\vec{\lambda}$. Let us remark that knowing the graph associated with $J$ does not mean knowing the values of the elements of $J$, but only which elements of $J$ are equal to $0$, i.e.\ knowing the relationships of influence among the processes. The matrix $t^*$ of the times of correlation must be known a priori in every case. 

In the context of operational risk real data come in the form of a database of historical operational losses; such a database is a collection of loss events occurred inside a bank; in order to be suitable for the estimation scheme that we are describing, the database must keep track of the amount, the process in which and the time at which each loss event occurred. The time resolution of the database is identified with the discrete time step of the model and the time at which the oldest loss occurred with $t = 0$, so that the database can be thought of as a realization of \eqref{eq:motion}. Since in this section there is no risk of ambiguity in the notation, the amount of loss registered in the database at the time step $t$ in the process $i$ will be denoted with $l_i(t)$.

\subsection{Estimating $\vec{\theta}$} \label{subsec:learning_theta}
In order to estimate $\theta_i$ let us look in the database of operational losses for the events such that $C_{ij}(t) = 0$, $\forall \, j$; assuming that the database is a realization of \eqref{eq:motion} we have:
\begin{equation} \label{eq:theta_1}
	l_i(t) = \Ramp \left[ \theta_i + \xi_i(t) \right] \, ;
\end{equation}
the probability that $l_i(t)=0$, conditioned on the occurrence on such events is:
\begin{equation} \label{eq:theta_2}
	\Pr \left[l_i(t)=0 \, | \, C_{ij}(t) = 0, \; \forall \, j \right] = \Pr \left[ \xi_i \leq - \theta_i \right] \, ,
\end{equation}
where the dependence of $\xi_i$ on $t$ has been dropped since its distribution does not depend on time. In order to make a frequentist estimate of the left hand side of \eqref{eq:theta_2} one would need a sample of values of $l_i(t)$, which is obviously not possible using a single database which contains only one value of $l_i$ at the time $t$; however, since the right hand side of \eqref{eq:theta_2} does not depend on time, also the left hand side must not:
\begin{equation} \label{eq:theta_3}
	\begin{split}
		\Pr \left[ l_i=0 \, | \, C_{ij} = 0, \; \forall \, j \right] & = \Pr \left[ \xi_i \leq - \theta_i \right] \\
		& = \int_{0}^{-\theta_i} {\lambda_i e^{-\lambda_i \xi_i} d\xi_i} \\
		& = 1 - e^{\lambda_i \theta_i} \, ,
	\end{split}
\end{equation}
where the left hand side has the meaning of a frequentist estimate from the database:
\begin{equation} \label{eq:theta_4}
	\Pr \left[ l_i=0 \, | \, C_{ij} = 0, \; \forall \, j \right] = \frac{\Fr \left[ (l_i=0), \; (C_{ij} = 0, \; \forall \, j) \right]}{\Fr \left[ C_{ij} = 0, \; \forall \, j \right]} \, .
\end{equation}
$\theta_i$ can be estimated inverting \eqref{eq:theta_3}:
\begin{equation} \label{eq:theta_5}
	\theta_i = \frac{1}{\lambda_i} \log \left( 1 - \Pr \left[ l_i=0 \, | \, C_{ij} = 0, \; \forall \, j \right] \right) \, ;
\end{equation}
let us explicitly notice from \eqref{eq:theta_5} that the values of $\theta_i$ estimated in such a way are negative. 

Let us make an example using the excerpt of a possible database shown in Tab.\ \ref{tab:data}; for simplicity we assume that $t_{ij}^* = 2$, $\forall \, i$ and $j$. Let us suppose to be interested in estimating the value of $\theta_1$; according to \eqref{eq:theta_4} we need to count the events such that $C_{1j} = 0$, $\forall \, j$; the counting starts from the first time step and proceeds using a moving window of width equal to $t_{1j}^*$ time steps: in this case one starts considering time steps $1-2$ and subsequently moves to $2-3$, $3-4$, etc. From Tab.\ \ref{tab:data} and using \eqref{eq:trigger} we see that for the event corresponding to time steps $1-2$ we have $C_{1j} = 0$, $\forall \, j$, meaning that $\Fr \left[ C_{1j} = 0, \; \forall \, j \right]$ must be incremented by one. To count the events such that $l_1 = 0$ \emph{and} $C_{1j} = 0$, $\forall \, j$ one has to consider one more time step: as $l_1(3) = 0$, also $\Fr \left[ (l_1=0), \; (C_{1j} = 0, \; \forall \, j) \right]$ must be incremented by one. 

\begin{table}[t]
	\centering
	\caption{Excerpt of six time steps of a possible database composed by five processes, where each row corresponds to a different time step and each column to a different process. Losses are indicated by $\bullet$, while empty spaces correspond to zero losses: e.g.\ both $l_3(3)$ and $l_3(4)$ are different from zero.}
	\label{tab:data}
	\vspace{0.5cm}
	\begin{tabularx}{0.75 \columnwidth}{CCCCCC}
		\toprule
			  & 1 & 2 & 3 & 4 & 5 \\
		\midrule
			1 &   &   &   &   &   \\
			2 &   &   &   &  &   \\
			3 &   &   & \bullet &   &   \\
			4 &   &   & \bullet &  &   \\
			5 &   & \bullet  &   & \bullet  &   \\
			6 & \bullet  &   &  &  &   \\			
			\ldots &   &   &   &   &   \\			
		\bottomrule
	\end{tabularx}
\end{table}

\subsection{Estimating $J$} \label{subsec:learning_gei}
The estimation of $J_{ij}$ uses the same line of reasoning followed to estimate $\theta_i$ from which differs only by the fact that it is based on different kinds of events; in this case we look for the events such that $C_{ij}(t) = c$ with $c=1, \ldots, t_{ij}^*$ and $C_{ik}(t) = 0, \; k \neq j$; for such events \eqref{eq:motion} reads:
\begin{equation} \label{eq:gei_1}
	l_i(t) = \Ramp \left[ c J_{ij} + \theta_i + \xi_i(t) \right] \, ;
\end{equation}
the probability that $l_i(t)=0$, conditioned on the occurrence on such events is:
\begin{equation} \label{eq:gei_2}
	\Pr \left[ l_i(t)=0 \, | \, C_{ij}(t) = c, \, C_{ik}(t) = 0, \; k \neq j \right]
	= \Pr \left[ \xi_i \leq - \theta_i - c J_{ij} \right] \, ;
\end{equation}
proceeding like in \eqref{eq:theta_3} we find:
\begin{equation} \label{eq:gei_3}
	\begin{split}
		\Pr \left[ l_i=0 \, | \, C_{ij} = c, \, C_{ik} = 0, \; k \neq j \right] & = \Pr \left[ \xi_i \leq - \theta_i - c J_{ij} \right] \\
		& = \int_{0}^{-\theta_i - c J_{ij}} {\lambda_i e^{-\lambda_i \xi_i} d\xi_i} \\
		& = 1 - e^{\lambda_i \left( \theta_i + c J_{ij} \right)} \, ,
	\end{split}
\end{equation}
where the left hand side of \eqref{eq:gei_3} has again the meaning of a frequentist estimate:
\begin{equation} \label{eq:gei_4}
	\Pr \left[ l_i=0 \, | \, C_{ij} = c, \, C_{ik} = 0, \; k \neq j \right] = \frac{\Fr \left[ (l_i=0), \; (C_{ij} = c, \, C_{ik} = 0, \; k \neq j) \right]}{\Fr \left[ C_{ij} = c, \, C_{ik} = 0, \; k \neq j \right]}
\end{equation}
and $J_{ij}$ can be estimated inverting \eqref{eq:gei_3}:
\begin{equation} \label{eq:gei_5}
	J_{ij} = \frac{1}{c} \left[ - \theta_i + \frac{1}{\lambda_i} \log \left( 1 - \Pr \left[ l_i=0 \, | \, C_{ij} = c, \, C_{ik} = 0, \; k \neq j \right] \right) \right] \, .
\end{equation}
Let us notice that \eqref{eq:gei_5} puts a subtle constraint on the parameters that can be estimated: $c J_{ij} + \theta_i < 0$, $\forall \, c$; if $\theta_i < 0$ (which is the case we are interested in) this translates into $t_{ij}^* J_{ij} < | \theta_i |$. 

In the context of operational risk the constraints imposed by \eqref{eq:theta_5} and \eqref{eq:gei_5} mean that the bank is exerting a control on the processes so strong that the interactions alone are not sufficient to generate a loss; in such a scenario a loss occurs when the noise is greater than the threshold set by the negative $\theta_i$ and the interaction term (if $J_{ij} > 0$) provides a mechanism to dynamically lower this threshold. In the case of a practical implementation, the soundness of these contraints should be certainly checked by experts in the organizational structure of the bank.

Also in this case Tab.\ \ref{tab:data} can be used to clarify how the events relative to the estimation of $J_{ij}$ are identified. Let us suppose to be interested in the estimation of $J_{43}$; from \eqref{eq:gei_4} we see that we need to count the events such that $C_{43} = c$, $C_{4k} = 0$, for $k \neq 3$; from Tab.\ \ref{tab:data} we see that time steps $2-3$ contribute to the case in which $c = 1$, while time steps $3-4$ contribute to the case in which $c = 2$, meaning that both $\Fr \left[ C_{43} = 1, \, C_{4k} = 0, \; k \neq 3 \right]$ and $\Fr \left[ C_{43} = 2, \, C_{4k} = 0, \; k \neq 3 \right]$ must be incremented by one. Since $l_4(4) = 0$, also $\Fr \left[ (l_4=0), \; (C_{43} = 1, \, C_{4k} = 0, \; k \neq 3) \right]$ is incremented by one, while it is not the case for $\Fr \left[ (l_4=0), \; (C_{43} = 2, \, C_{4k} = 0, \; k \neq 3) \right]$, since $l_4(5) \neq 0$.

\subsection{Estimating $\vec{\lambda}$} \label{subsec:learning_lambda}
In order to estimate the value of $\lambda_i$ the exact expression of $\langle l_i(t) \rangle$ will be exploited; since it is available only in the case in which the subgraph associated to the process $i$ has no loops, the discussion will be restricted to this case. If this is true for all the processes, the whole graph associated with the coupling matrix $J$ has no loops and $\lambda_i$ can be estimated $\forall \, i$. Let us start with the case of a free process $i$; using \eqref{eq:free_1}, \eqref{eq:theta_5} and \eqref{eq:zeta_analytical} we have:
\begin{equation} \label{eq:lambda_free}
	\lambda_i = \frac{T}{z_i(T)} \left( 1 - \Pr \left[ l_i=0 \, | \, C_{ij} = 0, \; \forall \, j \right] \right) \, ,
\end{equation}
where the case $\theta_i < 0$ of \eqref{eq:free_1} has been considered since \eqref{eq:theta_5} does not allow positive estimates of $\theta_i$. In \eqref{eq:lambda_free} $\langle z_i(T) \rangle$ has been replaced by the actual value calculated from the database of operational losses basing on the following argument; $z_i(t)/t$ is the sample average of the random variables $l_i(t)$ which are i.i.d.\ with finite mean given by \eqref{eq:free_1}; according to the law of large numbers $z_i(t)/t \rightarrow \langle l_i(t) \rangle$ that, together with \eqref{eq:zeta_analytical}, yields $z_i(t)/t \rightarrow \langle z_i(t) \rangle / t$; as discussed at the end of Section\ \ref{sec:solutions}, this argument only applies to all the cases in which the coupling matrix $J$ has no loops. 

For a process $i$ that is influenced only by a single free process $j$, \eqref{eq:one_leaf_4}, \eqref{eq:theta_5}, \eqref{eq:gei_5} and \eqref{eq:zeta_analytical} yield:
\begin{multline} \label{eq:lambda_one_leaf}
	\lambda_i = \frac{T}{z_i(T)} \sum_{c=0}^{t_{ij}^*} \left( 1 - \Pr \left[ l_i=0 \, | \, C_{ij} = c, \, C_{ik} = 0, \; k \neq j \right] \right) \\
	\cdot \binom{t_{ij}^*}{c} \left( 1 - \Pr \left[ l_i=0 \, | \, C_{ij} = 0, \; \forall \, j \right] \right)^c \\ 
	\cdot \left( \Pr \left[ l_i=0 \, | \, C_{ij} = 0, \; \forall \, j \right] \right)^{t_{ij}^* - \, c} \, ,
\end{multline}
where again the case $\theta_i < 0$ from \eqref{eq:free_1} has been considered and $\langle z_i(T) \rangle$ has been replaced by $z_i(T)$. Once $\lambda_i$ has been estimated through \eqref{eq:lambda_free} or \eqref{eq:lambda_one_leaf} and inserted into \eqref{eq:theta_5} and \eqref{eq:gei_5}, $\theta_i$ and $J_{ij}$ can be also estimated.

In the more general case in which the coupling matrix $J$ has no loops \eqref{eq:zeta_analytical} still applies and \eqref{eq:lambda_free} and \eqref{eq:lambda_one_leaf} can be extended using \eqref{eq:theta_5}, \eqref{eq:gei_5} and the results in the Appendix. In the most general case in which the matrix $J$ has causal loops, $\lambda_i$ may be elicited in an empirical way by assessing the mean value of a spontaneous loss in the process $i$, or by inverting \eqref{eq:free_3} and assessing the probability that a spontaneous loss occurs in the same process.

\section{Results} \label{sec:tests}
In order to check the consistency of the method proposed to estimate the parameters of the model we go after the following steps: i) we let the system evolve for $T$ time steps, ii) interpret the resulting trajectory (which will be called original trajectory in the following) as a database of operational losses and estimate the parameters, iii) insert the estimated parameters in \eqref{eq:motion} and sample a great number of trajectories, iv) compare $z_i^*(t)$, the cumulative loss of the original trajectory, with the average of $z_i(t)$ over the sample of trajectories. Since from \eqref{eq:gei_5} there may be up to $t_{ij}^*$ different estimates of $J_{ij}$ one may use the mean of the estimated $J_{ij}$ or sample from them. There are two reasons to perform the comparison basing on the cumulative losses $z_i(t)$ rather on $l_i(t)$: first, as already pointed out in Section\ \ref{sec:model}, $z_i(t)$ is the quantity of interest in the context of operational risk; second, at least in the case in which $J$ has no causal loops, $z_i(t)$ has the peculiar property to be self-averaging in time, i.e.\ $z_i(t) \rightarrow \langle z_i(t) \rangle$ (see Section\ \ref{subsec:learning_lambda}), being perfectly suitable to be compared with its average.

A slightly modified version of the previous strategy allows to test for the forecasting capability of the model as well: it is sufficient to estimate the parameters using only the first $f T$ (with $0 < f \leq 1$) time steps in the original trajectory, but still sampling trajectories lasting $T$ time steps; in this way we try to reproduce the behavior of $z_i(t)$ in the last $(1-f) T$ time steps ignoring the information contained in the same time steps of the original trajectory. For $f = 1$ the test on the forecasting capabilities reduces to the consistency check. In the case in which the matrix $J$ is known to have no causal loops it is not necessary to simulate the trajectories using \eqref{eq:motion}, but all the quantities of interest such as $\langle z_i(t) \rangle$ or $\var z_i(t)$ may be rather directly calculated by means of the exact solutions.

Let us briefly comment on the parameters chosen to generate the original trajectory. From \eqref{eq:motion} we see that $\theta_i$ may be chosen to be the unit of measurement of $l_i$ by properly rescaling $\theta_i$, $J_{ij}$ and the noise, so that one can take $\theta_i = \pm 1$, the sign being the same of $\theta_i$ before the rescaling; we are forced to choose $\theta_i = - 1$, $\forall \, i$ because \eqref{eq:theta_5} does not allow the estimation of positive $\theta_i$.

\begin{figure}[t]
	\centering
	\includegraphics[width = 0.2\columnwidth]{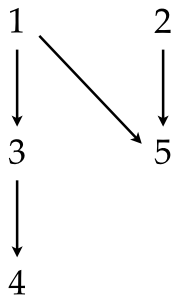}
	\caption{Graph associated with the matrix $J$. The nodes labeled $1$ and $2$ correspond to free processes; the process $3$ is influenced only by a free process (node $1$), while the process $5$ is influenced by two free processes (nodes $1$ and $2$); the process $4$ is influenced only by a process (node $3$) which is influenced only be a free process (node $1$).}
	\label{fig:graph}
\end{figure}

We stress that the number of processes ($N = 5$ in this case) does not play any significant role; actually the only relevant element is the complexity of the subgraphs contained into the graph associated with the matrix $J$: for this reason we start from the simplest subgraph (a free process) and move on considering progressively more complex ones. In fact, the structure of the matrix $J$ is chosen to encompass all the cases explicitly treated in the Appendix: free process ($i = 1, 2$), process influenced only by a free process ($i = 3$), process influenced only by a process which is influenced only by a free process ($i = 4$) and process influenced by two free processes ($i = 5$). The graph representing the influences among the processes is shown in Fig.\ \ref{fig:graph}: since it has no loops it is possible to estimate also $\vec{\lambda}$. In order to satisfy the constraint imposed by \eqref{eq:gei_5} we choose:
\begin{equation}
	J = 
	\begin{pmatrix}
		0 & 0 & 0 & 0 & 0 \\
		0 & 0 & 0 & 0 & 0 \\
		0.1 & 0 & 0 & 0 & 0 \\
		0 & 0 & 0.15 & 0 & 0 \\
		0.1 & 0.1 & 0 & 0 & 0 \\
	\end{pmatrix}
\end{equation}
and $t_{ij}^* = 5$, for $i$ and $j$ such that $J_{ij} \neq 0$. The  values $\lambda_i$ are chosen basing on the following argument; the more events suitable for the estimation of $\vec{\theta}$ and $J$ are found, the more the estimated values will be reliable; the events suitable for the estimation of $\vec{\theta}$ (see \eqref{eq:theta_2}) are more likely to be found in a database with a low density of losses, however, if this density becomes too low, there will be no events left to perform the estimation of $J$ (see \eqref{eq:gei_2}). We find that a reliable estimation of $\vec{\theta}$ and $J$ is obtained using:
\begin{equation}
	\vec{\lambda} = (2, 3, 5, 5, 5)
\end{equation}
and $T = 2 \cdot 10^5$. The initial condition used is: $l_i(t) = 0$, for $i = 1, \ldots, 5$, corresponding to a state in which all processes do not generate losses and thus can be considered perfectly functional.

For $f = 1$ the parameters are estimated with the following relative errors:
\begin{gather*}
	\delta \vec{\theta} \simeq (0.0033, \, 0.0029, \, 0.0390, \, 0.0074, \, 0.0343) \\
	\delta J_{31} \simeq 0.0959 \qquad \delta J_{43} \simeq 0.1313 \\
	\delta J_{51} \simeq 0.0377 \qquad \delta J_{52} \simeq 0.1466 \\
	\delta \vec{\lambda} \simeq (0.0030, \, 0.0032, \, 0.0407, \, 0.0022, \, 0.0337) \, ,
\end{gather*}
while for $f = 0.75$:
\begin{gather*}
	\delta \vec{\theta} \simeq (0.0044, \, 0.0032, \, 0.0468, \, 0.0094, \, 0.0369) \\
	\delta J_{31} \simeq 0.0659 \qquad \delta J_{43} \simeq 0.0009 \\
	\delta J_{51} \simeq 0.0566 \qquad \delta J_{52} \simeq 0.1520 \\
	\delta \vec{\lambda} \simeq (0.0033, \, 0.0052, \, 0.0445, \, 0.0012, \, 0.0332) \, .
\end{gather*}

\begin{figure*}[t]
	\centering
	\includegraphics[width = 0.48 \textwidth]{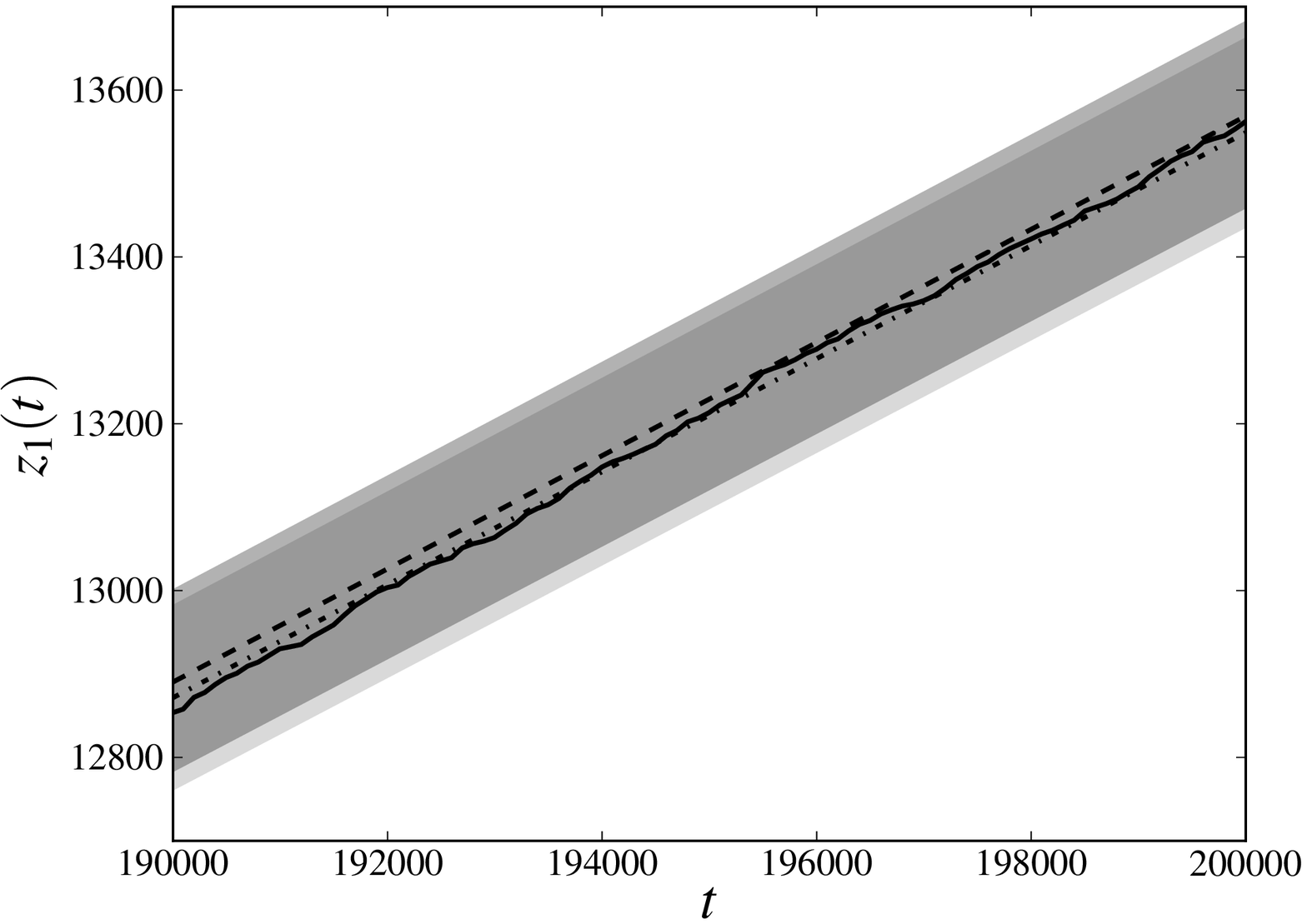}	
	\includegraphics[width = 0.48 \textwidth]{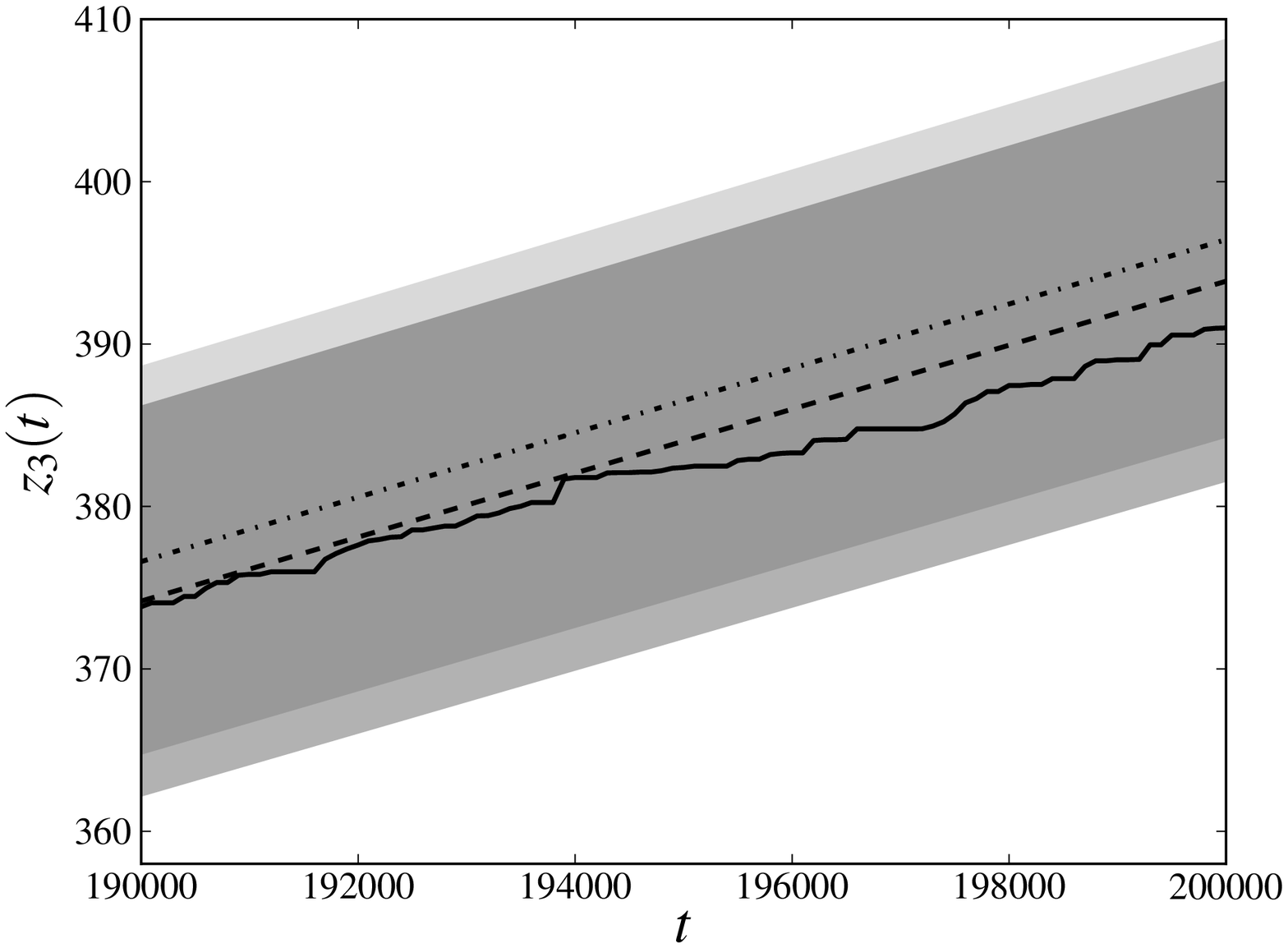}
	\includegraphics[width = 0.48 \textwidth]{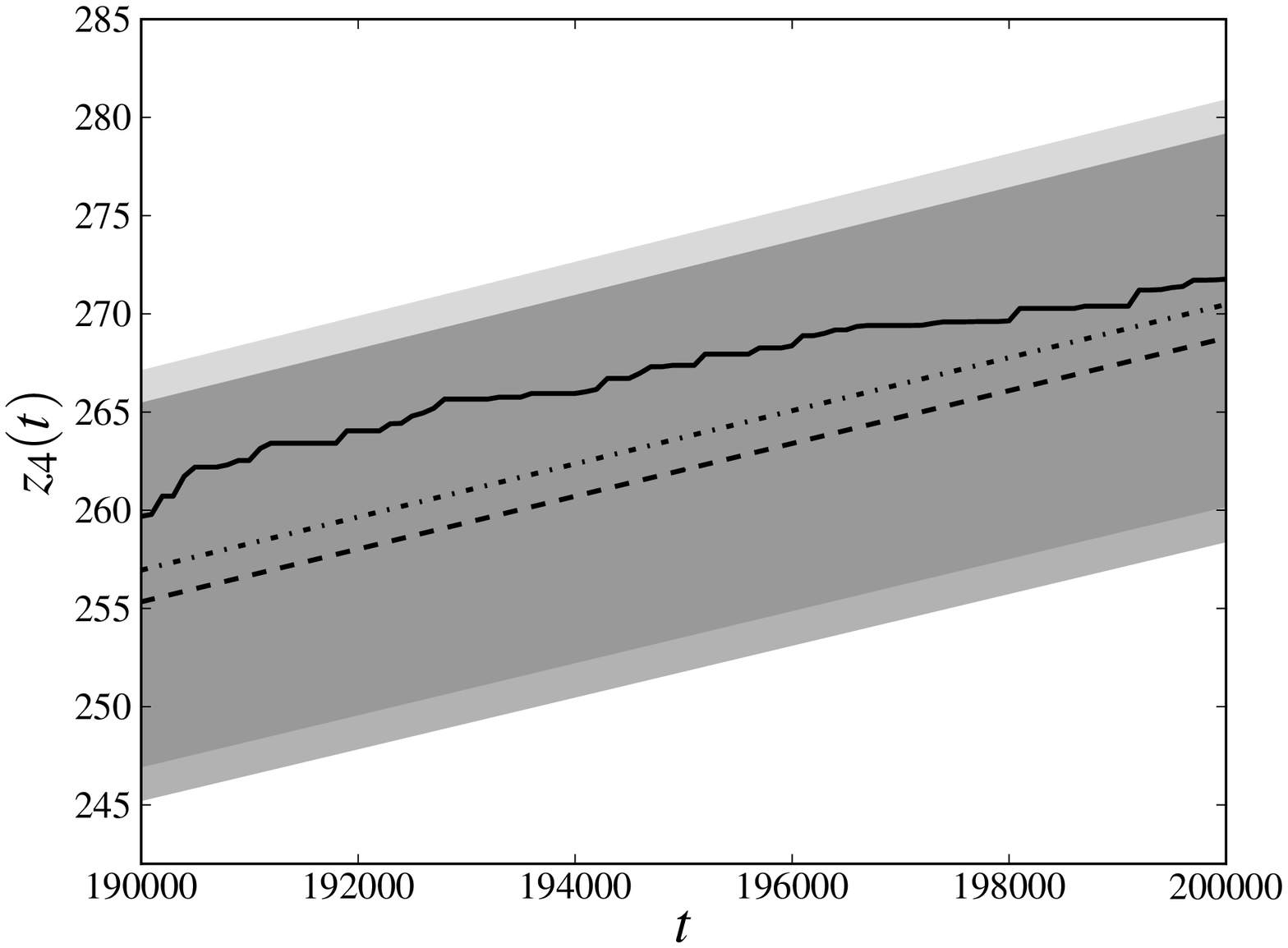}	
	\includegraphics[width = 0.48 \textwidth]{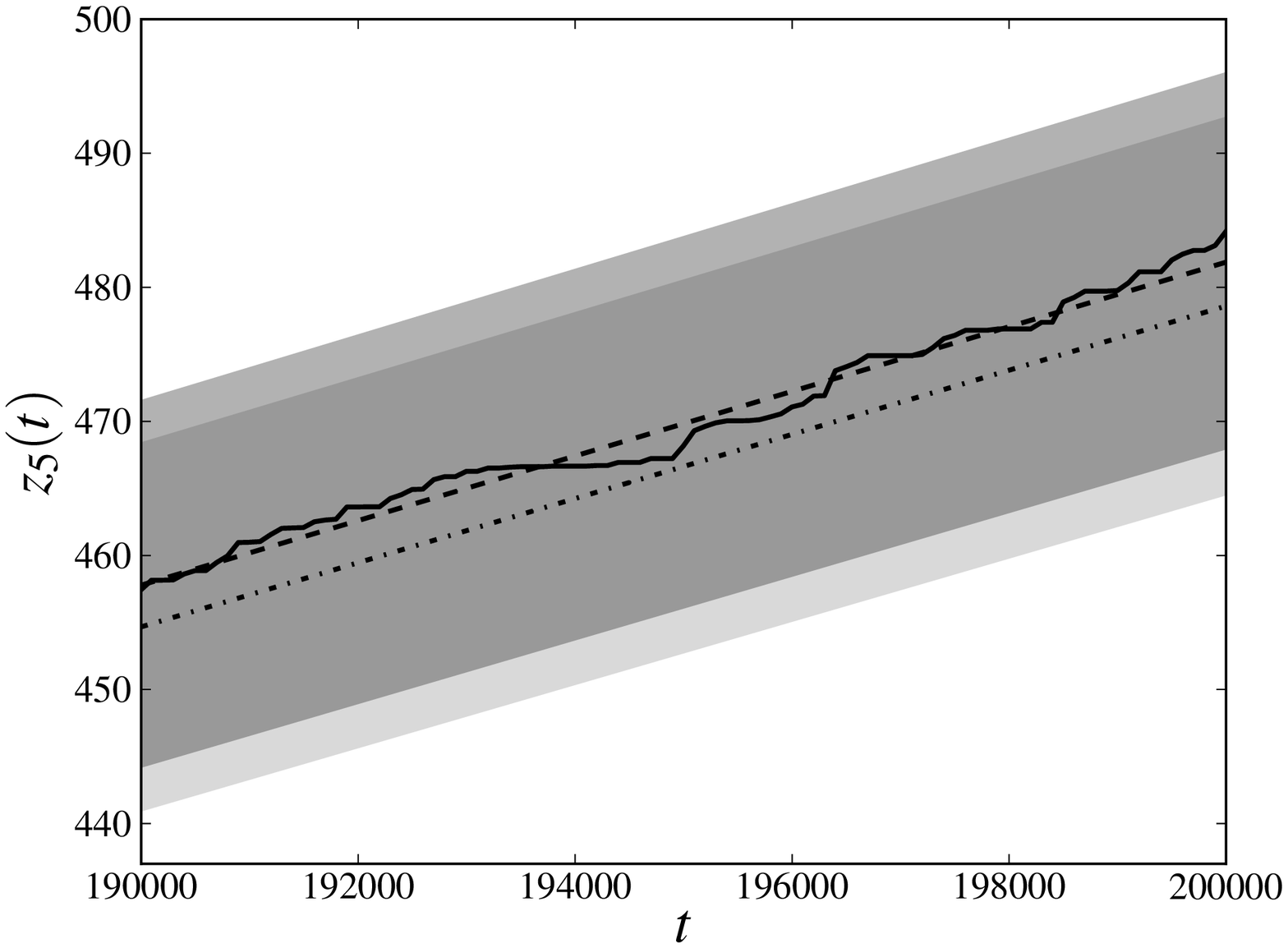}	
	\caption{$z_i^*(t)$, the cumulative loss of the original trajectory (solid line) and $\langle z_i(t) \rangle$, the average of $z_i(t)$ over the noise obtained estimating the parameters from the original trajectory, for $f = 1$ (dashed line) and $f = 0.75$ (dashed-dotted line); the limits of the dark (light) semi-transparent region are $\langle z_i(t) \rangle \pm \sigma_{z_i}(t)$ for $f = 1$ ($f = 0.75$); the darkest region is the overlap between the regions relative to $f = 1$ and $f = 0.75$. For all the processes $z_i^*(t)$ is reproduced with an uncertainty which is far less than $\sigma_{z_i}(t)$ and the error regions overlap almost completely.}
	\label{fig:cumul}
\end{figure*}

In Fig.\ \ref{fig:cumul} we compare $z_i^*(t)$, the cumulative loss of the original trajectory (solid line) with $\langle z_i(t) \rangle$, the average over the noise of $z_i(t)$ obtained estimating the parameters from the original trajectory and calculated with \eqref{eq:free_1}, \eqref{eq:one_leaf_4}, \eqref{eq:leaf_leaf_6} and \eqref{eq:two_leaf_3}, for $f = 1$ (dashed line) and $f = 0.75$ (dashed-dotted line); the semi-transparent regions span one standard deviation $\sigma_{z_i}(t) = \sqrt{\var z_i(t)}$ around $\langle z_i(t) \rangle$ and have been calculated by means of \eqref{eq:free_2}, \eqref{eq:one_leaf_5} and the analogues of \eqref{eq:leaf_leaf_6} and \eqref{eq:two_leaf_3} for the variance. Since both the process $i = 1$ and the process $i = 2$ are free and their results are qualitatively identical, we only show those relative to the process $i = 1$; only the last $10^4$ time steps are shown for the sake of readability. The fact that $z_i^*(t)$ is reproduced for all the processes with an error which is far less than one standard deviation for $f = 1$ proves the consistency of the estimation of the parameters proposed in Section\ \ref{sec:learning}; the same result for $f = 0.75$ shows that the model exhibits the capability to forecast the cumulative losses in the last quarter of the original trajectory. Moreover, the error regions relative to $f = 1$ and $f = 0.75$ overlap almost completely for all the processes: this means that all the relevant information about the parameters of the model is contained in the fraction of the database used for the estimation and that the information contained in the remaining part is redundant. 

\begin{figure}
	\centering
	\includegraphics[width = 0.6 \textwidth]{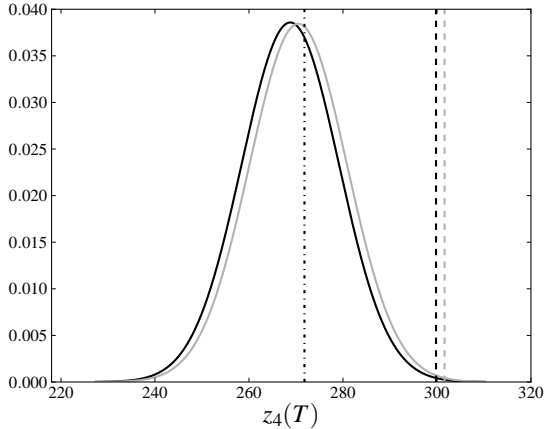}	
	\caption{$z_4^*(T)$, the cumulative loss of the original trajectory at the final time step (dashed-dotted line) and the Gaussian distribution of $z_4(T)$ obtained estimating the parameters from the original trajectory, for $f = 1$ (solid dark line) and $f = 0.75$ (solid light line). The two distributions overlap almost completely and their peaks correspond to $z_4^*(T)$. The relative error of the VaRs over the time horizon $T$ and with level of confidence $99.865$ for $f = 1$ (dashed dark line) and $f = 0.75$ (dashed light line) is $\simeq 10^{-3}$.}
	\label{fig:dist}
\end{figure}

In Fig.\ \ref{fig:dist} we show $z_4^*(T)$ (dashed-dotted line) and the Gaussian distribution of $z_4(T)$ obtained estimating the parameters from the original trajectory, for $f = 1$ (solid dark line) and $f = 0.75$ (solid light line). Fig.\ \ref{fig:dist} refers to the process $i = 4$ since its associated subgraph is the more complex; the results obtained for the other processes are completely analogous. We notice that the two distributions overlap almost completely and that their peaks correspond to $z_4^*(T)$. 

The VaR over the time horizon $T$ and with level of confidence $99.865$ can be easily calculated for a Gaussian distribution,  being equal to $\langle z_i(t) \rangle + 3 \, \sigma_{z_i}(t)$; in Fig.\ \ref{fig:dist} the VaRs of the process $4$ for $f = 1$ (dashed dark line) and $f = 0.75$ (dashed light line) are shown to be almost identical: their relative error is $< 10^{-3}$. In Tab.\ \ref{tab:var} the VaRs are reported for $f = 1$ and $f = 0.75$, together with their relative error $\delta \text{VaR}$ which is $\simeq 10^{-3}$ for all the processes.

\begin{table}[t]
	\centering
	\caption{VaRs over the time horizon $T$ and with level of confidence $99.865$ for the process $i$ calculated from the cumulative losses $z_i(T)$ obtained estimating the parameters from the original trajectory, for $f = 1$ and $f = 0.75$; $\delta \text{VaR}$, the relative error between $\text{VaR}^{f=1}$ and $\text{VaR}^{f=0.75}$ is $\simeq 10^{-3}$ for all the processes.}
	\label{tab:var}
	\vspace{0.5cm}
	\begin{tabularx}{0.75 \columnwidth}{CCCC}
		\toprule
		i & \text{VaR}^{f=1} & \text{VaR}^{f=0.75} & \delta \text{VaR} \\
		\midrule
		1 & 13 \, 906.79 & 13 \, 886.83 & 1.43 \cdot 10^{-3} \\
		2 &  3 \, 337.36 &  3 \, 360.42 & 6.88 \cdot 10^{-3} \\
		3 &       430.68 &       433.29 & 6.05 \cdot 10^{-3} \\ 
		4 &       299.79 &       301.61 & 6.03 \cdot 10^{-3} \\
		5 &       524.13 &       520.70 & 6.56 \cdot 10^{-3} \\
		\bottomrule
	\end{tabularx}
\end{table}

As pointed out in Section\ \ref{sec:solutions}, if the variance of the noise is not finite, the distribution of $z_i(t)$ for large $t$ is not Gaussian; using the generalized limit theorem \cite{gnedenko-kolmogorov}, it is possible to show that the distribution of $z_i(t)$ must be positively skewed and heavy-tailed, in agreement with some empirical results \cite{moscadelli}. However, the analysis of the model with infinite variance of the noise poses both conceptual and computational problems: since the variance of $l_i(t)$ diverges as well, another reliable measure of the width of the distribution must be found to quantify the goodness of the model predictions; the relationship between the parameters of the distribution of $l_i(t)$ and the parameters of the cumulative loss distribution $z_i(t)$ is entirely different from \eqref{eq:zeta_analytical} and must be found; the rate of convergence to the non-Gaussian limit distributions is much slower and even a purely numerical analysis is much more computationally expensive.

\section{Comparison with an alternative model} \label{sec:alt_model}
Let us now consider an alternative model with a slightly different dynamics:
\begin{equation} \label{eq:motion2}
	\hat{l}_i(t) = s_i(t) \cdot n_i(t) \, ,
\end{equation}
where $s_i(t)$ is drawn from some distribution, which we are leaving unspecified for the moment, and independently from the noise, while
\begin{equation} \label{eq:n}
	n_i(t) = \Theta \left( \sum_{j=1}^N J_{ij} C_{ij}(t) + \theta_i + \xi_i(t) \right) \, .
\end{equation}
This dynamics involves only the variables $n_i(t)$, while the time dependence in $s_i(t)$ should be intended has a mere label to distinguish different values drawn from the same distribution. $\hat{l}_i(t)$ has still the meaning of the loss of the process $i$ at the time $t$ and, according to \eqref{eq:motion2}, is nonzero if $n_i(t)$ is equal to one, i.e.\ if the argument of the Heaviside function in \eqref{eq:n} is larger than zero, which is the same condition under which $l_i(t)$ in \eqref{eq:motion} is nonzero. The dynamics of $n_i(t)$ controls the number of time steps in which the losses occur, while the amount of the losses depends on the distribution of $s_i(t)$. From this point of view, such a dynamics model can be considered a generalization of the LDA, and it makes sense to call frequency the stochastic process associated to $n_i(t)$ and severity the random variable associated to $s_i(t)$. Models based on a dynamics similar to the one defined by \eqref{eq:motion2} and \eqref{eq:n} have been introduced in \cite{kuhn-neu-2003, anand-kuhn}. Since frequency and severity are independent, one can imagine to fix the severity distribution so that the dynamics of \eqref{eq:motion2} is equivalent to the one of \eqref{eq:motion} \emph{on average}, in the sense that $l_i(t)$ and $\hat{l}_i(t)$ have the same distribution; it can be done by choosing the severity so that:
\begin{equation} \label{eq:s_constraint}
	\Pr[s_i(t) > x] = \frac{\Pr[l_i(t) > x]}{\Pr[n_i(t) = 1]} = \frac{\Pr[l_i(t) > x]}{\langle n_i(t) \rangle} \, ,
\end{equation}
where the r.h.s.\ can be calculated once that all the parameters ($\vec{\theta}$, $J$, and $\vec{\lambda}$) are known. As a consequence, the dynamics defined by \eqref{eq:motion2} and \eqref{eq:n} appears more general than the one defined by \eqref{eq:motion}, since, at least in principle, it permits to use arbitrary severity distributions, even the ones not satisfying \eqref{eq:s_constraint}. However, using an arbitrary severity distribution in general does not allow to employ the procedure described in Section\ \ref{subsec:learning_lambda} to estimate the parameters of the noise.

Actually it is possible to show that fixing the mean of the severity so that:
\begin{equation} \label{eq:s_mean}
	\langle s_i(t) \rangle = \frac{\langle l_i(t) \rangle}{\langle n_i(t) \rangle} \, ,
\end{equation}
is sufficient to estimate $\lambda_i$. For a free process $i$ and for $\theta_i < 0$ one has that:
\begin{equation} \label{eq:s_lambda_free}
	\begin{split}
		\langle s_i(t) \rangle & = \frac{\langle z_i(t) \rangle}{\langle n_i(t) \rangle \, t} \, = \frac{\langle z_i(t) \rangle}{p_i^F(\theta_i) \, t} \\
		& = \frac{z_i(t)}{t} \cdot \frac{1}{\left( 1 - \Pr \left[ l_i=0 \, | \, C_{ij} = 0, \; \forall \, j \right] \right)} = \frac{1}{\lambda_i} \, ,
	\end{split}
\end{equation}
where \eqref{eq:zeta_analytical} has been used, and the last equalities derive respectively from the self-averaging property of the cumulative loss and from \eqref{eq:lambda_free}.\footnote{In the case of a noise distribution different from the one in \eqref{eq:noise} one has to verify that, like in \eqref{eq:s_lambda_free}, $\langle s_i(t) \rangle$ depends only on the parameters of the noise.} With a little thought this result can be easily extended to the more general case in which the matrix $J$ has no causal loops.\footnote{The calculation of $\langle n_i(t) \rangle$ is similar to the one of $\langle l_i(t) \rangle$ that has been carried out in Section\ \ref{sec:solutions} and in the Appendix.} Nevertheless, there is no solid argument that justifies to impose the constraint \eqref{eq:s_mean} on the severity. In particular, it is not clear why the mean should be the only moment of the severity depending on the parameters of the model (more specifically of the noise). Therefore, it would be desirable to have a coherent model that allows to derive \eqref{eq:s_mean}, and thus to estimate the parameters of the noise. The crucial observation is that the constraint \eqref{eq:s_constraint}, and consequently the model whose dynamics is defined by \eqref{eq:motion}, implies the constraint \eqref{eq:s_mean}. Hence, the model whose dynamics is defined by \eqref{eq:motion} has the virtue that it allows to estimate the parameters of the noise in a coherent way, both in the sense that the constraint \eqref{eq:s_mean} has not to be imposed ``by hand'' and in the sense that all the terms appearing in the equation of motion have a clear interpretation (explicitly corresponding to a mechanism for producing or avoiding operational losses, as explained in Section\ \ref{sec:model}). Moreover, it is worth to point out that, in the case in which \eqref{eq:s_constraint} holds, the severity has a much clearer relationship with the parameters of the model; it is straightforward to show\footnote{Using \eqref{eq:s_constraint} and noting that the distribution of $l_i(t)$ can be calculated observing that $\Pr[l_i(t) > x] = \int_0^\infty \Theta[l_i(t) - x] d\tilde{\xi}_i(t)$.} that for a free process $i$ and for $\theta_i < 0$:
\begin{equation} \label{eq:s_excess}
		 F_{s_i(t)}(x) = \frac{F_{\xi_i(t)}(x - \theta_i) - F_{\xi_i(t)}(- \theta_i)}{1 - F_{\xi_i(t)}(- \theta_i)} \, ,
\end{equation}
i.e.\ that the severity has the excess distribution of the noise over the threshold $\theta_i$ (with $F_y(x) = \Pr[y \leq x]$ we denote the distribution of the random variable $y$, evaluated in $x$). As regards the more general case in which the matrix $J$ has no causal loops, it is still possible to show that a generalized version of \eqref{eq:s_excess} holds, where both the numerator and the denominator are replaced by linear combinations whose coefficients depends on the topology of the graph associated to the matrix $J$, similarly to the cases treated in the Appendix.

\section{Conclusions} \label{sec:outro}
In this paper we proposed a dynamical model to forecast operational losses in banks. The equation of motion provides two different mechanisms for the generation of losses in a process: the interaction with other processes and the spontaneous generation due to a random noise; since the different-time correlations play a crucial role in this context, the interactions are non-local in time; the effort made by the bank to avoid the occurrence of losses is also taken into account by means of an inhomogeneous external field. We have shown that, if the coupling matrix $J$ is known to have no causal loops, all the parameters of the model except the maximum times of correlations $t_{ij}^*$ can be estimated from real data, so that the model can be tailored on the internal organizational structure of a specific bank; in the most general case also the parameters of the noise must be known a priori. Focusing on the case in which the coupling matrix $J$ is known to have no causal loops, we exactly solve the model and find the asymptotic behaviour of the cumulative loss, showing that the non-locality of the equation of motion is not sufficient alone to modify the shape of the cumulative loss distribution. We specialize the procedure for estimating $\vec{\theta}$ and $J$ suggested in \cite{anand-kuhn} to the considered model, propose a procedure to estimate the parameters of the noise, and validate it.

Many statistical approaches, like the static LDA, are founded on the implicit hypothesis that the basic statistical properties of the distributions of operational losses do not change in time; basing on this assumption the capital charge that the bank has to put aside to face operational risk the \emph{next} year is calculated from the loss distribution built from historical data. The assumption made by the approach proposed here and in \cite{kuhn-neu-2003} is definitely weaker and consists in assuming that the basic mechanisms underlying the generation of operational losses do not change in time. The crucial advantage of such an approach is that it allows to make forecasts about future losses. The forecasting power of the model has been investigated estimating the parameters of the model only from a fraction $f$ of a simulated database of operational losses and comparing the cumulative losses of the remaining part with those forecast by the model. We have shown that the model exhibits surprisingly good capabilities in forecasting the future losses even for $f = 0.75$: in particular the relative error between the actual VaR ($f = 1$) and the forecast VaR ($f = 0.75$) is $\simeq 10^{-3}$ for all the processes. In order to check the performances of the proposed model, both the validation of the parameters estimation and the test of the forecasting power has been carried out using simulated data.

We think that the general framework of dynamical models for operational risk deserves further investigation in several directions; let us just cite few examples: the case in which the coupling matrix has causal loops could be explored, more complex terms of interaction in the equation of motion could be considered or different mechanisms for the generation of losses included; as explained in Section\ \ref{sec:tests}, the study of the case in which the variance of distribution of the noise is not finite looks particularly promising as it may lead to the emergence of heavy-tailed cumulative loss distributions.

\appendix
\section{}
The results \eqref{eq:free_1}, \eqref{eq:free_2}, \eqref{eq:one_leaf_4} and \eqref{eq:one_leaf_5} will be extended in two particular cases. In the first case the process $i$ is influenced only by the process $j$, which in turn is influenced only by the process $k$ which is free ($i \leftarrow j \leftarrow k$). In this case the average over the noise is:
\begin{equation}	\label{eq:leaf_leaf_1}
	\langle l_i(t) \rangle = \int_0^{\infty} l_i(t) \prod_{1 \leq s \leq t_{ij}^*} \tilde{d \xi_j}(t-s) \, \tilde{d \xi_i}(t)
	\prod_{2 \leq r \leq t_{ij}^* + t_{jk}^*} \tilde{d \xi_k}(t-r) \, ;
\end{equation}
the events $C_{ij}(t)=0, \ldots, C_{ij}(t)=t_{ij}^*$ still cover the entire domain of integration, but are not mutually exclusive: in fact $C_{ij}(t)$ depends through $l_j(t-1), l_j(t-2), \ldots, l_j(t-t_{ij}^*)$ on $C_{jk}(t-1), C_{jk}(t-2), \ldots, C_{jk}(t-t_{ij}^*)$ which in turn have crossed dependences from $l_k(t-2), l_k(t-3), \ldots, l_k(t- t_{ij}^* - t_{jk}^*)$ so that, for example, both $C_{jk}(t-1)$ and $C_{jk}(t-2)$ depend on $l_{k}(t-3)$. However, it is still possible to rewrite \eqref{eq:leaf_leaf_1} in the following way:
\begin{equation} \label{eq:leaf_leaf_2}
	\begin{split}
		\langle l_i(t) \rangle = & \sum_{\{c\} } \int_0^{\infty} \Ramp \left( J_{ij} \sum_{s'=1}^{t_{ij}^*}c_{s'} + \theta_i + \xi_i(t) \right) \tilde{d \xi_i}(t) \\
		& \cdot \int_{\{ \Theta \left[ l_j(t-s'') \right] = c_{s''}\}_{s''}} \prod_{1 \leq s \leq t_{ij}^*} \tilde{d \xi_j}(t-s) \, 
	\prod_{2 \leq r \leq t_{ij}^* + t_{jk}^*} \tilde{d \xi_k}(t-r) = \\
		= & \sum_{\{c\} } m_i^F \Big( J_{ij} \sum_{s'=1}^{t_{ij}^*}c_{s'} + \theta_i \Big) \\
		& \cdot \int_{\{ \Theta \left[ l_j(t-s'') \right] = c_{s''}\}_{s''}} \prod_{1 \leq s \leq t_{ij}^*} \tilde{d \xi_j}(t-s) \, 
	\prod_{2 \leq r \leq t_{ij}^* + t_{jk}^*} \tilde{d \xi_k}(t-r) \, ,
	\end{split}
\end{equation}
where the sum over $\{c\}$ is over all the possible  configurations $c_1 \in \{0, 1\}, \ldots,$ $c_{t_{ij}^*} \in \{0, 1\}$. Once a particular configuration $\{c\}$ has been assigned, the integral on the right hand side of \eqref{eq:leaf_leaf_2} is simply the probability that $\Theta \left[ l_j(t-s') \right] = c_{s''}$, for $s''= 1,\ldots,t_{ij}^*$ and equals to:
\begin{multline}	\label{eq:leaf_leaf_3}
	 \int_{\Big\{ \Theta \left[ J_{jk} + \sum_{r'=1}^{t_{jk}^*}\Theta \left[ l_k(t-s''-r') \right] + \theta_j + \xi_j(t-s'') \right] = c_{s''} \Big\}_{s''}} \prod_{1 \leq s \leq t_{ij}^*} \tilde{d \xi_j}(t-s) \, \\
	\cdot \prod_{2 \leq r \leq t_{ij}^* + t_{jk}^*} \tilde{d \xi_k}(t-r) = \\
	= \sum_{\{ d \}} \int_{\Big\{ \Theta \left[ J_{jk} + \sum_{r''=s''}^{s''+t_{jk}^*} d_{r''} + \theta_j + \xi_j(t-s'') \right] = c_{s''} \Big\}_{s''}} \; \prod_{1 \leq s \leq t_{ij}^*} \tilde{d \xi_j}(t-s) \\
	\cdot \int_{\Big\{ \Theta \left[ l_k(t-r') \right] = d_{r'} \Big\}_{r'}} \prod_{2 \leq r \leq t_{ij}^* + t_{jk}^*} \tilde{d \xi_k}(t-r) \, ,
\end{multline}
where again the sum over $\{d\}$ is analogous to the sum over $\{c\}$ and $r'= 2,\ldots, t_{ij}^* + t_{jk}^*$. We notice that integrals on the right hand side of \eqref{eq:leaf_leaf_3} are decoupled and can be respectively rewritten as:
\begin{multline}	\label{eq:leaf_leaf_4}
	\prod_{1 \leq s \leq t_{ij}^*} \int_{ \Theta \left[ J_{jk} + \sum_{r'=s}^{s+t_{jk}^*} d_{r'} + \theta_j + \xi_j(t-s) \right] = c_ {s}} \tilde{d \xi_j}(t-s) = \\
	= \prod_{1 \leq s \leq t_{ij}^*} \left[ p_j^F \Big(J_{jk} \sum_{r'=s}^{s+t_{jk}^*} d_{r'} + \theta_j\Big)  \delta_{c_s,1} +\Big[ 1 - p_j^F \Big(J_{jk} \sum_{r'=s}^{s+t_{jk}^*} d_{r'} + \theta_j\Big) \Big] \delta_{c_s,0} \right] \, ,
\end{multline}
\begin{equation}	\label{eq:leaf_leaf_5}
	\prod_{2 \leq r \leq t_{ij}^* + t_{jk}^*} \int_{\Theta \left[ l_k(t-r) \right] = d_{r}} \tilde{d \xi_k}(t-r) = \prod_{2 \leq r \leq t_{ij}^* +  t_{jk}^*} \left[ p_j^F \left( \theta_k \right)  \delta_{c_r,1} + \Big[ 1 - p_j^F \left( \theta_k\right) \Big] \delta_{c_r,0} \right] \, .
\end{equation}
Using \eqref{eq:leaf_leaf_2}, \eqref{eq:leaf_leaf_3}, \eqref{eq:leaf_leaf_4} and \eqref{eq:leaf_leaf_5} one finally obtains:
\begin{multline}	\label{eq:leaf_leaf_6}
	\langle l_i(t) \rangle = \sum_{\{c\} } m_i^F \Big( J_{ij} \sum_{s'=1}^{t_{ij}^*}c_{s'} + \theta_i \Big) \\
	\cdot \sum_{\{d\}} \prod_{1 \leq s \leq t_{ij}^*} \left[ p_j^F \Big(J_{jk} \sum_{r'=s}^{s+t_{jk}^*} d_{r'} + \theta_j\Big)  \delta_{c_s,1} + \Big[ 1 - p_j^F \Big(J_{jk} \sum_{r'=s}^{s+t_{jk}^*} d_{r'} + \theta_j\Big) \Big] \delta_{c_s,0} \right] \\
	\cdot \prod_{2 \leq r \leq t_{ij}^* + t_{jk}^*} \left[ p_j^F \left( \theta_k \right)  \delta_{c_r,1} + \Big[ 1 - p_j^F \left( \theta_k\right) \Big] \delta_{c_r,0} \right] \, ,
\end{multline}
while the variance is easily obtained from \eqref{eq:leaf_leaf_6} by replacing $m_i^F$ with $m_i^{(2)F}$ and subtracting $\langle l_i(t) \rangle^2$. The value of $\lambda_i$ can again be estimated from \eqref{eq:free_3}, \eqref{eq:leaf_leaf_6}, \eqref{eq:theta_5} and \eqref{eq:gei_5}, analogously to \eqref{eq:lambda_one_leaf}. This case can be trivially extended to all the graphs which are simple paths and contain $m$ nodes, i.e.\ to all the graphs of the type $i_1 \leftarrow i_2 \leftarrow \ldots \leftarrow i_{m-1} \leftarrow i_{m}$. We point out that in \eqref{eq:leaf_leaf_6} one has to sum $2^{t_{ij}^* + t_{jk}^*}$ terms; in the case in which the simplest loop were present in the graph associated with the matrix $J$, i.e.\ a loop of the process $i$ with itself ($J_{ii} \neq 0$), it is easy to argue that the number of terms to sum in order to calculate $\langle l_i(t) \rangle$ would be equal to $2^{t t_{ii}^*}$; in fact, such a topology is equivalent to a simple path containing $t$ copies of the process $i$.

In the second case that we will consider the process $i$ is influenced only by two processes $j_1$ and $j_2$ that are both free. In this case $l_i(t)$ depends only on $l_{j_1}(t-1), \ldots, l_{j_1}(t-t_{ij_1}^*)$ through $C_{ij_1}(t)$ and on $l_{j_2}(t-1), \ldots, l_{j_2}(t-t_{ij_2}^*)$ through $C_{ij_2}(t)$, so that the average over the noise equals to the average over the random variables $\xi_i(t), \, \xi_{j_1}(t-1), \ldots, \xi_{j_1}(t-t_{ij_1}^*), \, \xi_{j_2}(t-1), \ldots, \xi_{j_2}(t-t_{ij_2}^*)$ and \eqref{eq:one_leaf_1} and \eqref{eq:one_leaf_2} read:
\begin{multline} \label{eq:two_leaf_1}
	\langle l_i(t) \rangle = \int_0^{\infty} \Ramp \left[ J_{ij_1} C_{ij_1}(t) + J_{ij_2} C_{ij_2}(t) + \theta_i + \xi_i(t) \right] \\
	\cdot \prod_{1 \leq s_1 \leq t_{ij_1}^*} \tilde{d \xi_{j_1}}(t-s_1) \prod_{1 \leq s_2 \leq t_{ij_2}^*} \tilde{d \xi_{j_2}}(t-s_2) \; \tilde{d \xi_i}(t) \, ,
\end{multline}
\begin{multline} \label{eq:two_leaf_2}
	\prod_{1 \leq s \leq t_{i{j_1}}^*} \int_0^{\infty} \tilde{d \xi_{j_1}}(t-s) \, \prod_{1 \leq s \leq t_{i{j_2}}^*} \int_0^{\infty} \tilde{d \xi_{j_2}}(t-s) = \\
	= \left( \sum_{c_1=0}^{t_{i{j_1}}^*} \int_{C_{ij_1}(t) = c_1} \prod_{1 \leq s_1 \leq t_{ij_1}^*} \tilde{d \xi_{j_2}}(t-s_1) \right)
	\cdot \left( \sum_{c_2=0}^{t_{i{j_2}}^*} \int_{C_{ij_2}(t) = c_2}  \prod_{1 \leq s_2 \leq t_{ij_2}^*} \tilde{d \xi_{j_2}}(t-s_2) \right) \, ,
\end{multline}
where the domain of integration of the variables $\xi_{j_1}(t-1), \ldots, \xi_{j_1}(t-t_{ij_1}^*), \, \xi_{j_2}(t-1), \ldots, \xi_{j_2}(t-t_{ij_2}^*)$ has been divided in subsets with fixed values of $C_{ij_1}(t)$ and $C_{ij_2}(t)$. Inserting \eqref{eq:two_leaf_2} and \eqref{eq:one_leaf_3} into \eqref{eq:two_leaf_1} one obtains:
\begin{equation} \label{eq:two_leaf_3}
	\begin{split}
		\langle l_i(t) \rangle = & \sum_{c_1=0}^{t_{i{j_1}}^*}  \binom{t_{i{j_1}}^*}{c_1} \left[ p_{j_1}^F(\theta_{j_1}) \right]^{c_1} \left[ 1 - p_{j_1}^F(\theta_{j_1}) \right]^{t_{ij_1}^* - \, c_1} \\
		& \cdot \sum_{c_2=0}^{t_{i{j_2}}^*}  \binom{t_{i{j_2}}^*}{c_2} \left[ p_{j_2}^F(\theta_{j_2}) \right]^{c_2} \left[ 1 - p_{j_2}^F(\theta_{j_2}) \right]^{t_{ij_2}^* - \, c_2} \\
		& \cdot \, m_i^F(c_1 J_{i{j_1}} + c_2 J_{i{j_2}} + \theta_i) \, .
	\end{split}
\end{equation}
As in the aforementioned case, the variance is obtained from \eqref{eq:two_leaf_3} by replacing $m_i^F$ with $m_i^{(2)F}$ and subtracting $\langle l_i(t) \rangle^2$. Analogously to \eqref{eq:lambda_one_leaf}, the value of $\lambda_i$ can be estimated from \eqref{eq:free_3}, \eqref{eq:two_leaf_3}, \eqref{eq:theta_5} and \eqref{eq:gei_5} and the constraint $t_{ij_1}^* J_{ij_1} + t_{ij_2}^* J_{ij_2} < | \theta_i |$ emerges. This case can be also trivially extended to all the graphs in which the process $i$ is influenced by an arbitrary number of (say $m$) free processes, leading to the general constraint $t_{ij_1}^* J_{ij_1} + \ldots +t_{ij_m}^* J_{ij_m} < | \theta_i |$.

In the more general case in which the graph representing the interactions has no loops both $\langle l_i(t) \rangle$ and $\var l_i(t)$ are sums over all the simple paths starting from a leaf node and ending to the node $i$ which can be calculated combining the extensions to the first and second case treated in the Appendix. Also in this general case  both $\langle l_i(t) \rangle$ and $\var l_i(t)$ do not depend on time and are finite, allowing to extend the results \eqref{eq:lambda_free} and \eqref{eq:lambda_one_leaf} of Section\ \ref{subsec:learning_lambda}.

\section*{Acknowledgments}
M.~B. would like to thank Maria Valentina Carlucci for the countless suggestions and useful discussions.

\end{document}